\documentclass{article}

\usepackage{arxiv}

\usepackage[utf8]{inputenc} 
\usepackage[T1]{fontenc}    
\usepackage{hyperref}       
\usepackage{url}            
\usepackage{booktabs}       
\usepackage{amsfonts}       
\usepackage{nicefrac}       
\usepackage{microtype}      
\usepackage{graphicx}
\usepackage{doi}
\usepackage{subfig}
\usepackage{color, soul}
\usepackage{authblk}
\usepackage{float}
\usepackage{calc}
\usepackage{indentfirst}
\usepackage{fancyhdr}
\usepackage{graphicx,epstopdf}
\usepackage{lastpage}
\usepackage{ifthen}
\usepackage[right]{lineno}
\usepackage{comment}
\usepackage{amsmath}
\usepackage{amssymb} 
\usepackage{setspace}
\usepackage{enumitem}
\usepackage{mathpazo}
\usepackage{titlesec}
\usepackage{etoolbox} 
\usepackage{tabto} 
\usepackage{xcolor, colortbl} 
\usepackage{soul} 

\usepackage{multirow}
\usepackage{microtype} 
\usepackage{tikz} 
\usepackage{totcount} 
\usepackage{changepage} 
\usepackage{attrib} 
\usepackage{upgreek} 
\usepackage{array} 
\usepackage{tabularx}
\usepackage{pbox} 

\usepackage[english]{babel}
\usepackage{fancyhdr}

\title{Dataset of Multi-aspect Integrated Migration Indicators}

\author{Diletta Goglia$^{a}$$^{1}$, Laura Pollacci$^{a}$$^{2}$, Alina S\^irbu$^{a}$$^{3}$ \\
        \small $^{a}$Department of Computer Science, University of Pisa, Pisa, Italy \\
        \small $^{1}$\texttt{d.goglia@studenti.unipi.it} (D.G.)\\
        \small $^{2}$\texttt{laura.pollacci@di.unipi.it} (L.P.)\\
        \small $^{3}$\texttt{alina.sirbu@unipi.it} (A.S.)
}

\hypersetup{
pdftitle={Dataset of Multi-aspect Integrated Migration Indicators},
pdfauthor={Diletta Goglia, Laura Pollacci, Alina Sîrbu},
pdfkeywords={International migration, migration nowcasting, migration flows, migration stocks, migration drivers, bilateral migration, social connectedness index, Facebook social connectedness.},
}

\date{}

\begin{document}
\maketitle

\begin{abstract}
	Nowadays, new branches of research are proposing the use of non-traditional data sources for the study of migration trends in order to find an original methodology to answer open questions about the human mobility framework.
    In this context we presents the Multi-aspect Integrated Migration Indicators (MIMI) dataset, an new dataset of migration drivers, resulting from the process of acquisition, transformation and merge of both official data about international flows and stocks and original indicators not typically used in migration studies, such as online social networks. 
    This work describes the process of gathering, embedding and merging traditional and novel features, resulting in this new multidisciplinary dataset that we believe could significantly contribute to nowcast to forecast both present and future bilateral migration trends.
\end{abstract}


\vspace{1cm}

\section{Introduction} \label{section:intro}

In the last years the pursuit of original drivers and measures is becoming an increasing requirement to migration studies, considering the new methods and technologies used to characterize and understand human migration phenomenon.
Many researchers \cite{sirbu2021, JRC127369, alexander2022, delfava2019} have proposed to employ non-traditional data sources to study migration trends, including so-called social Big Data such as online social networks. The usefulness of exploiting unconventional data sources for better understanding migration patterns, as well as the benefits of merging knowledge from both traditional and novel datasets, have already been proven \cite{sirbu2021}.
This unconventional approach is intended to find an alternative methodology to ultimately answer open questions about the human mobility framework (i.e. nowcasting flows and stocks, studying integration of multiple sources and knowledge, and investigating migration drivers).
Nevertheless, in this context of meaningful combination of the conventional and the original, many types of data exist, still very scattered and heterogeneous: in the variety of this background, integration is not straightforward. 

For this purpose we propose a tool to be exploited in migration studies as a concrete example of this new integration-oriented approach: the Multi-aspect Integrated Migration Indicators (MIMI) dataset. It includes both official data about bidirectional human mobility (traditional flow and stock data) with multidisciplinary features and original indicators, including the Facebook Social Connectedness Index (SCI), which measures the relative probability that two individuals across two countries are friends with each other on Facebook. The inclusion of SCI in the dataset enables it to be exploited as a non-traditional way to describe, understand and nowcast international migration. The combination of this index with socioeconomic variables measuring the similarity of two locations (such as per capita income, religiosity and language) already appeared in \cite{Bailey2018, bailey2020determinants} where it has been shown that pairs of locations that are more similar on these dimensions share more friendship links. Nevertheless a similar approach on country level is still missing; moreover, such observations and conclusions about SCI have never been exploited in migration studies.\\ For this reason, our aim is to use this "homophily" concept (defined as the empirical regularity with which individuals are more likely to be associated with other individuals of similar characteristic) \cite{Lazarsfeld1954}, that literature has already linked to Facebook social connectedness \cite{Bailey2018, bailey2020determinants}, to present a new dataset useful for better understanding country-to-country human mobility trends. \\

\section{Motivation} \label{section:motivation}
MIMI is an open dataset that provides multidimensional information about several traditional and non-traditional aspects related to human mobility phenomenon. 
Thanks to this variety of knowledge, experts from several research fields (demographers, sociologists, economists) could exploit MIMI to investigate the behavior of many drivers and relate it to migration trends, so as to build a comprehensive overview and understanding of them.\\
As an example, it could be possible to access existing correlations between original sources of data and traditional migration measures, explore and investigate them and try to identify any possible causal relationship.\\
Moreover, it could be possible to develop complex models able to assess human mobility framework by evaluating related interdisciplinary drivers, as well as models able to nowcast and predict traditional migration indicators in accordance with original features, such as the strength of social connectivity. By means of these algorithms, companies and researchers could find an alternative methodology to  answer open questions about emerging mobility trends.\\

Human migration is a complex phenomenon characterized by several related factors. 
It is also ancient as human history, and it has been widely studied, explored and described over time. However, the technological advancements and the rapid and drastic changes that society faced in the 21st century have impacted on the human mobility phenomenon, which consequently has undergone radical modifications. We believe that taking into account this same information about society changes and technological progress (such as economic, cultural and social big data) can be an effective strategy nowadays to detect new trends in bilateral migration and to better understand and nowcast it.\\
The motivations for building and releasing the MIMI dataset precisely lie in this need of new perspectives, methods and analyses that can no longer prescind from taking into account a variety of new factors. The heterogeneous and multidimensional sets of data present in MIMI offer an all-encompassing overview of the characteristics of international human mobility, enabling a better understanding and an original potential exploration of the relationships between migration and non-traditional sources of data.


\section{Data description} \label{section:data} 
The MIMI \cite{goglia2022} dataset version 1 (March 15, 2022) was released under the Creative Commons Attribution 4.0 International Public License (CC BY 4.0\footnote{\url{https://creativecommons.org/licenses/by-nc/4.0/}}) and is publicly available on Zenodo  (\href{https://doi.org/10.5281/zenodo.6360651}{10.5281/zenodo.6360651}). It consists of a single file containing more than 28,000 entries (records) and 480 different features.\\ In this section we provide all the dataset specifications and describe the structure of the CSV file in detail, as well as how each feature was built.

\subsection{Data structure} 
\subsubsection{Data files and format}
The MIMI dataset is made up of one single CSV file that includes 28,725 rows and 485 columns. The index consists in uniquely identified pairs of countries, built from the join of the two ISO-3166 alpha-2 codes of origin and destination country respectively. Indeed, the dataset contains as main features country-to-country bilateral migration flows and stocks, together with the Facebook strength of connectedness of each pair. 

\subsubsection{Geographical coverage}
The dataset comprises migration features and social strenght of Facebook connectedness for 254 different countries belonging to the following macro-areas: North America, South America, Europe, Asia, Africa, Oceania, Antarctica.

\subsubsection{Temporal coverage}
Since our work does not focus on the study of migration phenomenon \textit{per sé} but on its possible relationship with social networks, in particular with the use of Facebook, the choice of the time range has been calculated accordingly.
Therefore, the initial decision was not to select migration data antecedent to 2004.\\
However, our intention was to make available a tool that could also be useful for the study on the differences between contemporary and past trends (e.g. alterations of some phenomenons, consistent changes of values compared to the past, consequences of previous data on the last few years, etc...): for this reason some features have been selected starting from 2000.\\
Certainly, data selection according to predetermined temporal ranges always depends on the availability of sources: for example, during our data collection phase, Eurostat was not providing information about population density of countries before 2008.\\
Table \ref{table:temp_coverage} provides a detailed temporal coverage of each time-related feature, apart from SCI for which we included the only one made available (the latest, which refers to October 13, 2021, updated in December 15, 2021).

\begin{table}[H] 
\caption{Temporal coverage of each time-related features. "End" always refers to the latest available measure. For all the abbreviations refer to Section \ref{abbr}. \label{table:temp_coverage}}

\newcolumntype{C}{>{\centering\arraybackslash}X}
\begin{tabularx}{\textwidth}{llp{12cm}}
\toprule
\multicolumn{3}{l}{\textit{Yearly measures}} \\
\toprule
\textbf{Start}	& \textbf{End}	& \textbf{Feature}\\
\midrule
2000	    & 2020			& GDP at PPP\\
2000	    & 2022			& UN population\\
2008		& 2019			& EUROSTAT population density, EUROSTAT total immigrants and emigrants for each country\\
2010		& 2019			& EUROSTAT migration flows\\
2010		& 2020			& UN migration flows, UN total immigrants and emigrants for each country\\
2010		& 2021			& EUROSTAT population\\
\toprule
\multicolumn{3}{l}{\textit{Five-year measures}} \\
\toprule
\textbf{Start}	& \textbf{End}	& \textbf{Feature}\\
\toprule
2000	    & 2020			& Migration stocks\\
2000		& 2025			& NET migration and NET migration rate\\

\bottomrule
\end{tabularx}
\end{table}
\unskip


\subsection{Dataset features} 

\subsubsection{Features definition} \label{section:features_def} 
In this section we are going to list all the indicators included in the MIMI dataset, then we will describe them in detail in the following section.\\
Table \ref{table:features} contains a complete declaration of all drivers, grouped and categorized by context ("feature area").\\
The column "Name" contains the identifier of each feature: since it would not be possible to list all features, a more compact replacement rule is presented in order to include them all in the table. From this simple rule it is possible to derive the exact name of each single indicator.
The column "Name" should be read as follows: the \textit{invariant} part of the identifier is static, while the \texttt{interchangeable} part must be substituted as explained below in order to obtain the exact name of the feature.

\begin{itemize}
    \item \texttt{country} should be replaced with \textit{origin} or \textit{destination}.
    \item \texttt{year} and \texttt{start-end} should be replaced, respectively, with the reference year (in case of annual feature) or reference year range (for NET migration and NET migration rate features). Substituted values should be consistent with the temporal coverage available for each indicator, which can be found in Table \ref{table:temp_coverage}.
    \item \texttt{source} allows \textit{UN} and \textit{ESTAT} as replacement values.
    \item \texttt{sex} should be substituted with \textit{F}, \textit{M} or \textit{T} (respectively, female, male or both).
    \item \texttt{age} allows only \textit{T} as replacement value for data obtained from UN (both flows and stocks), while it can take four different values for ESTAT flows: \textit{T} (total), \textit{<15} (less than 15 years), \textit{15-64} (from 15 to 64 years), \textit{>65} (65 years or over). 

\end{itemize}

Some examples are provided in Table \ref{table:features} footnotes. 

\begin{table}
\small
\caption{Features list. The exact name of the each single indicator can be retrieved by following the rule explained in Section \ref{section:features_def}.
\label{table:features}}
    \begin{adjustwidth}{-0,5cm}{0cm}
	    \definecolor{Gray}{gray}{0.85}
		\begin{tabularx}{1.1\textwidth}{p{2,5cm} >{\columncolor{Gray}}p{0,3cm} p{6cm} p{1,5cm} p{4,6cm}  p{1cm}}\\
		\toprule
			\textbf{Feature area}	& \cellcolor{white}  & \textbf{Name}	    & \textbf{Related to}     & \textbf{Brief description}    & \textbf{Dtype}\\
			\midrule
			
			Index \cellcolor[HTML]{f2f2f2}  & 1 & \textit{from\_to}  & \cellcolor[HTML]{f2f2f2} & Unique identifier of each record & object\\
			
			Facebook \cellcolor[HTML]{e6f2ff} & 2 & \textit{sci} & \cellcolor[HTML]{f2f2f2} \cellcolor[HTML]{f2f2f2} & Social Connectedness Index & float64\\
			
			\cellcolor[HTML]{f2f2f2} & 3  & \textit{from\_to\_cont} & \cellcolor[HTML]{f2f2f2}  & Pair of continents codes & object\\
			
			\cellcolor[HTML]{f2f2f2} & 4  & \textit{geodesic\_distance\_km}  &  \multirow{-4}{*}{\shortstack[l]{Pair of \\ countries}} \cellcolor[HTML]{f2f2f2} & Distance between countries      & float64\\
			
		    \cellcolor[HTML]{f2f2f2} & 5 & \texttt{country}\textit{\_country} & \cellcolor[HTML]{e6f2ff} &  ISO-3166-1 alpha-2 code & object \\
		    
			\cellcolor[HTML]{f2f2f2}  &6 & \texttt{country}\textit{\_name}&\cellcolor[HTML]{e6f2ff} & ISO-3166 name& object\\
			
			\cellcolor[HTML]{f2f2f2}  &7 & \texttt{country}\textit{\_alpha\_3} &\cellcolor[HTML]{e6f2ff}  & ISO-3166-1 alpha-3 code & object\\
			
			\cellcolor[HTML]{f2f2f2} &8 & \texttt{country}\textit{\_official\_name} &\cellcolor[HTML]{e6f2ff} & ISO-3166 official name & object\\
			
			\cellcolor[HTML]{f2f2f2} &9 & \texttt{country}\textit{\_cont\_code}&  \cellcolor[HTML]{e6f2ff}  & Continent code & object\\
			
			\cellcolor[HTML]{f2f2f2} &10 & \texttt{country}\textit{\_cont\_name} & \cellcolor[HTML]{e6f2ff} & Continent name & object\\
			
			\cellcolor[HTML]{f2f2f2} &11 & \texttt{country}\textit{\_latitude} & \cellcolor[HTML]{e6f2ff}  & Centroid latitude     & float64\\
			 \cellcolor[HTML]{f2f2f2} &12  & \texttt{country}\textit{\_longitude} & \cellcolor[HTML]{e6f2ff}& Centroid longitude & float64\\
			 
			\cellcolor[HTML]{f2f2f2} &13  & \texttt{country}\textit{\_coordinate} & \cellcolor[HTML]{e6f2ff} & Centroid \textit{(lat, long)} pair   & object\\
			
			\cellcolor[HTML]{f2f2f2} &14& \texttt{country}\textit{\_neighbors} &  \cellcolor[HTML]{e6f2ff} & List of bordering countries & object\\
			
			\multirow{-12}{*}{Geographic} \cellcolor[HTML]{f2f2f2} &15 & \texttt{country}\textit{\_area} & \cellcolor[HTML]{e6f2ff} & Area (in squared kilometers)& float64\\
                
			\cellcolor[HTML]{e6f2ff} &16  & \texttt{country}\textit{\_religion}& \cellcolor[HTML]{e6f2ff}  & List of religions & object \\
			
	        \cellcolor[HTML]{e6f2ff} &17 & \texttt{country}\textit{\_gdp}\texttt{\_year} &\cellcolor[HTML]{e6f2ff} & Annual GDP at PPP & float64 \\
	         
	        \cellcolor[HTML]{e6f2ff}  &18  & \texttt{country}\textit{\_languages} & \cellcolor[HTML]{e6f2ff} & List of spoken languages & object \\
	        
	        \cellcolor[HTML]{e6f2ff} &19  & \texttt{country}\textit{\_fb\_users}  & \cellcolor[HTML]{e6f2ff}  & Number of Facebook users & Int64 \\
	        
	        \cellcolor[HTML]{e6f2ff} &20  & \texttt{country}\textit{\_fb\_users\_perc} & \cellcolor[HTML]{e6f2ff} & Percentage of Facebook users & float64 \\
	        
	        \cellcolor[HTML]{e6f2ff} &21 & \texttt{country}\textit{\_PDI} &\cellcolor[HTML]{e6f2ff}  & \cellcolor[HTML]{f2f2f2} & Int64 \\
	        
	        \cellcolor[HTML]{e6f2ff} &22 & \texttt{country}\textit{\_IDV}  & \cellcolor[HTML]{e6f2ff} & \cellcolor[HTML]{f2f2f2}  & Int64 \\
	        
	        \cellcolor[HTML]{e6f2ff} &23  & \texttt{country}\textit{\_MAS} & \cellcolor[HTML]{e6f2ff} & \cellcolor[HTML]{f2f2f2} & Int64 \\
	        
	        \cellcolor[HTML]{e6f2ff} &24 & \texttt{country}\textit{\_UAI} & \cellcolor[HTML]{e6f2ff}  & \cellcolor[HTML]{f2f2f2} & Int64 \\
	        
			\multirow{-11}{*}{Interdisciplinary}\cellcolor[HTML]{e6f2ff} &25 & \texttt{country}\textit{\_LTO} & \cellcolor[HTML]{e6f2ff} & \multirow{-5}{*}{Cultural Indicators} \cellcolor[HTML]{f2f2f2} & Int64 \\
			 
            \cellcolor[HTML]{f2f2f2}  &26  & \texttt{source\_country\_}\textit{pop}\_\texttt{year}\textsuperscript{1} & \cellcolor[HTML]{e6f2ff} & Annual \textbf{population stocks} & Int64 \\
            
            \cellcolor[HTML]{f2f2f2}  &27 & \texttt{source\_country\_}\textit{pop\_dens}\_\texttt{year} &  \cellcolor[HTML]{e6f2ff}  & Annual \textbf{population density} & float64 \\
                
			\cellcolor[HTML]{f2f2f2} &28 & \texttt{source\_country\_}\textit{total\_imm}\_\texttt{year} & \cellcolor[HTML]{e6f2ff} & Annual \textbf{total immigrants} & Int64\\
			
			\cellcolor[HTML]{f2f2f2} &29 & \texttt{source\_country\_}\textit{total\_em}\_\texttt{year} & \cellcolor[HTML]{e6f2ff} & Annual \textbf{total emigrants} & Int64 \\
			
			\cellcolor[HTML]{f2f2f2} &30 & \texttt{source\_country\_}\textit{net\_migr}\_\texttt{start-end}\textsuperscript{2} & \cellcolor[HTML]{e6f2ff} & Five-year \textbf{NET migration} & float64\\
			
			\cellcolor[HTML]{f2f2f2} &31 & \texttt{source\_country\_}\textit{net\_migr\_rate}\_\texttt{start-end} & \multirow{-28}{*}{\shortstack[l]{Single\\country}}\cellcolor[HTML]{e6f2ff} & Five-year \textbf{NET migration rate} & float64\\

			\cellcolor[HTML]{f2f2f2} &32 & \texttt{source\_year\_sex\_age}\textsuperscript{3} & \cellcolor[HTML]{f2f2f2} & Annual \textbf{migration flows} & Int64\\
			
			\multirow{-8}{*}{Demographic }\cellcolor[HTML]{f2f2f2} &33 & \texttt{source\_}\textit{migr\_stocks}\texttt{\_year\_sex\_age}\textsuperscript{4} & \multirow{-2}{*}{\shortstack[l]{Pair of\\countries}} \cellcolor[HTML]{f2f2f2} & Five-year \textbf{migration stocks} & Int64\\
	
			\bottomrule
		\end{tabularx}\end{adjustwidth}
	
	\noindent{\footnotesize{
                \textsuperscript{1} e.g. \textit{ESTAT\_origin\_pop\_2017}. \\
                \textsuperscript{2} e.g. \textit{UN\_destination\_NET\_migr\_2005-2010}. \\
                \textsuperscript{3} e.g. \textit{ESTAT\_2015\_M\_>65}. \\
                \textsuperscript{4} e.g. \textit{UN\_migr\_stocks\_2000\_F\_T}. \\
                }
            }
\end{table}

\subsubsection{Features description and sources} \label{section:features_descr} 
In this section we are going to describe in detail each single feature listed in the previous section, also reporting all the data sources: some indicators may have multiple sources since they were necessary to better integrate missing values.\\
As stated in Section \ref{section:motivation}, the purpose of the integration of all these different drivers in the MIMI dataset is to allow the exploration of any of their possible connections with the international migration phenomenon, and eventually exploit them to better understand and nowcast it.

\begin{itemize} 
    \item \textbf{Index} (\textit{feature 1} in Table \ref{table:features}). \\ The index consists in \textbf{uniquely identified pairs of countries}, built as follows: \texttt{ISO2 code of origin country} \textbf{-} \texttt{ISO2 code of destination country} (e.g. AL-FI index indicates records related to migration from Albania to Finland). Pairs having the same country codes for origin and destination indicate the so-called "returners"
    (e.g. BE-BE record represents people that were born or have citizenship in Belgium which moved their residence in Belgium in the reference year). 
    \\
    
    \item \textbf{Facebook data} (\textit{feature 2} in Table \ref{table:features}). \\ This indicator represents one of the most non-traditional feature (i.e. social media data) within the context of migration studies that we included. It consists in the so-called Facebook \textbf{Social Connectedness Index} (\href{https://bit.ly/Facebook_SCI}{bit.ly/Facebook\_SCI})
    publicly provided by "Data for Good at Meta"\footnote{\url{https://bit.ly/DataForGoodAtMeta}} organisation on "Humanitarian Data Exchange, Data for Good" platform\footnote{\url{https://data.humdata.org/}} . Country-to-country values of SCI are available in TSV format for more that 34,000 pairs, updated to December 2021 \cite{sci2021}. \\ 
    
    This indicator uses anonymized insights of active Facebook users and their friendship networks to measure the intensity of connectedness between locations \cite{Bailey2018}. In this way, the resulting formulation in Equation \ref{equation} is a measure of the social connectedness between the two locations \textit{i} and \textit{j}, that is representative of the relative probability that two individuals across the two locations are friends with each other on Facebook: if $Social Connectedness Index_{i,j}$ is twice as large, a Facebook user in country \textit{i} is about twice as likely to be connected with a given Facebook user in country \textit{j}. 
    
    \begin{linenomath}
    \begin{equation} \label{equation}
    Social Connectedness Index_{i,j} = \frac{FB\_Connections_{i,j}}{FB\_Users_{i} \ast FB\_Users_{j}}
    \end{equation}
    \end{linenomath}
    
    Specifically, in this work the concept of "locations" coincides with NUTS0 areas since our dataset only focuses on country-to-country bilateral migration.  Nevertheless, SCI is also provided with respect to narrower geographical granularities, (e.g. NUTS2, NUTS3): we do not exclude future works focused on the study of migration trends at a smaller resolution (country-to-county, or county-to-county).\\
    
    The SCI has a symmetric structure by definition of the concept of "friendship" and has been re-scaled to have a maximum value of 1,000,000,000 and a minimum value of 1.
    In our dataset, the minimum possible value was originally 0 (indicating pairs of countries for which the index was not available), subsequently replaced with an arbitrarily small value (chosen as half of the minimum available) in order to fix problems when computing Pearson correlation of the logarithmic SCI.\\
    
    \item \textbf{Geographic features} (\textit{features 3-15} in Table \ref{table:features}). \\ These features portray and contextualize both origin and destination countries at geographical level providing all the necessary information to describe them, starting from the official codes and names, up to their land extent and how far they are.
    Specifically: \\
    \begin{itemize}
        \item \textit{features 5, 6, 7, 8}  are \textbf{ISO-3166 standards nomenclatures for country identification}, retrieved from PyCountry Python module\footnote{\url{https://pypi.org/project/pycountry/}} and ISAN  (International Standard Audiovisual Number) \cite{ISAN2021}.\\
        
        \item \textit{features 9, 10} identify \textbf{continents} as follows: Africa (AF); Antarctica (AQ); Asia (AS); Europe (EU); America, North (NA); Oceania (OC); America, South (SA).\\
        
        \item \textit{feature 3} consists in the pair \texttt{ code of origin continent} \textbf{ - } \texttt{code of destination continent}. Its functionality can be fully appreciated in chord diagrams of Section \ref{section:results}.\\
        
        \item \textit{features 11, 12, 13} \textbf{locate the position} of the centroids of both origin and destination countries in a classic geographic coordinate system. They are gathered and integrated from Google DSPL \cite{DSPL} and from \texttt{latlng()} method of CountryInfo Python library \footnote{\url{https://pypi.org/project/countryinfo/\#latlng}}, and then merged together in a tuple (\textit{feature 13}) built as a specific GeoPandas data structure called "geometry array"\footnote{\url{https://geopandas.org/en/stable/docs/reference/api/geopandas.points_from_xy.html}}.\\
        
        \item \textit{feature 4} is the \textbf{measure of distance} between origin and destination, computed starting from the tuple in \textit{feature 13} of both countries and using the geodesic formulation\footnote{\url{https://geopy.readthedocs.io/en/stable/\#module-geopy.distance}} \cite{Karney2013} provided by GeoPy Python library\footnote{\url{https://pypi.org/project/geopy/}}.\\ It has already been observed in \cite{Bailey2018} that, at county level, much of the estimated effect of distance on migration might be coming from the relationship between distance and social connectedness: therefore the use SCI indicator could better explain the variation of migration flows than geographic distance alone can. \\
        
        \item \textit{feature 14} consists in the list of countries that \textbf{share a border} with the given country. The utility of this feature is to find out if the two countries of origin and destination share a border, using a straightforward function to check if a country name (\textit{feature 6}) is contained into the list of neighbors of the other, and vice versa. An additional binary feature (e.g. "neighbors", having value \texttt{True} or \texttt{False}) could be derived from this method. Countries having empty list are islands. The corresponding sources for this feature are the following: GitHub repository in \cite{github2017}, \texttt{borders()} method of CountryInfo Python module\footnote{\url{https://pypi.org/project/countryinfo/\#borders}} and Wikipedia \cite{wiki:List_of_countries}.\\
        
        \item \textit{feature 15} is the \textbf{measure of the area} extension of the country in squared kilometers. It is gathered from The World Bank \cite{twb_area} and integrated with \texttt{area()} method of CountryInfo Python module\footnote{\url{https://pypi.org/project/countryinfo/\#area}}.\\
    \end{itemize}

    \item \textbf{Interdisciplinary indicators} (\textit{features 16-25} in Table \ref{table:features}). \\ Some of these drivers are considered non-traditional data in the context of migration studies since their use in migration understanding and nowcasting is poorly documented in literature.
    Despite this, most of the available studies consider these features as relevant in such context, as they are related to the behavior of international migration trends. \\ 
    \begin{itemize}
        \item \textit{feature 17} is an indicator that provides per capita\footnote{calculated as the aggregate of production (GDP) divided by the population size.} annual values for \textbf{gross domestic product (GDP)} of a country, expressed in current international dollars and converted by purchasing power parity (PPP)\footnote{a detailed definition of PPP provided by System of National Accounts 1993 Glossary can be found here: \url{https://unstats.un.org/unsd/nationalaccount/glossresults.asp?gID=438}} conversion factor. Data is retrieved from The World Bank \cite{twb_gdp}. 
        The gross domestic product is one of the "Development Indicators", already widely used in literature in combination with global migration.\\
        
        \item \textit{features 16, 18} correspond to two lists containing, respectively, the most practiced \textbf{religions}, and the most spoken \textbf{languages} in the country (both including official ones and minorities). 
        The benefit of including these columns would be to discover if the two countries of origin and destination share some languages or religions (or both), since this could favor a migratory exchange between the two.
        Rare languages and religions used only in one country and not shared with any other have been removed as meaningless for our purposes.\\ Languages have been gathered from Wikipedia \cite{wiki:lang} while religions comes from DataHub \cite{datahub_religion} and have been integrated with Wikipedia data \cite{wiki:religion}.\\
        
        \item \textit{features 19, 20} indicates the quantity (respectively, as absolute number and as percentage of the total population) of \textbf{Facebook users} that a given country has. The source is World Population review \cite{wpr}, which refers to the latest available measure for each country (oldest date back to December 2020).\\
        
        \item \textit{features 21-25} represents \textbf{Cultural Indices} of a location, intended as dimensions along which cultural values of that location can be analyzed \cite{Kaasa2014}. Their origin dates back to the work of \cite{Hofstede1980} although, over the decades, independent research branches led to the creation and addition of new ones \cite{wiki:Hofstede's_cultural_dimensions_theory}. Our work includes five of these indicators, of which we provide a brief individual description. \\ Their applications in literature have been several (e.g. cross-cultural studies using Twitter data \cite{ieee2019}), but the purpose of their inclusion in the MIMI dataset 
        is to use them in an original way: our intention is to explore and understand their possible relation with international migration trends.\\
        Data about cultural indicators are available in different NUTS levels but in our work they only appear related to NUT0 (country) level since it is the only one that fits our geographic viewpoint. 
        \\ \textit{Features 21-25} are the result of the integration of the two different datasets \cite{cultdim1, cultdim2}. Unfortunately, they are provided only for 66 of the more than 250 available countries but, despite this, most of them have already shown to be strongly involved in migration trends (see the behavior of their correlation values with the absolute number of migrants of a country, in Section \ref{section:corr}).\\
        Starting from cultural dimensions of both countries of origin destination, a new feature about cultural distance could be obtained: datasets with this configuration already exist
        \cite{Kaasa2014, KAASA2016231} despite, at the moment, data is available only for a third of the countries (22 in total).
        \\ 
        
        \begin{itemize}
            \item \textit{feature 21} is \textbf{Power distance indicator (PDI)} which is defined as “the extent to which the less powerful members of organizations and institutions (like the family) accept and expect that power is distributed unequally” \cite{wiki:Hofstede's_cultural_dimensions_theory}. This index describes the extent to which hierarchical relations and unequal distribution of power in organisations and societal institutions are accepted in a culture. \\
            
            \item \textit{feature 22}: \textbf{Individualism indicator (IDV)}\footnote{the same indicator can be found in other sources and contexts with the "IND" acronym which, however, can be confused with the Indulgence cultural indicator (IND).}
            (as opposed to collectivism) explores the “degree to which people in a society are integrated into groups” \cite{wiki:Hofstede's_cultural_dimensions_theory}: it reflects the extent to which people prefer to act as individuals rather than as members of a community.\\
            
            \item \textit{feature 23} is \textbf{Masculinity indicator (MAS)}, defined as “a preference in society for achievement, heroism, assertiveness and material rewards for success” \cite{wiki:Hofstede's_cultural_dimensions_theory}: as opposed to femininity, this dimension reveals to what degree traditionally masculine societal values, such as orientation towards accomplishment, prevail over values such as modesty, solidarity or tolerance. \\
            
            \item \textit{feature 24} is \textbf{Uncertainty avoidance indicator (UAI)} defined as “a society's tolerance for ambiguity”, in which people embrace or avert an event of something unexpected, unknown, or away from the status quo \cite{wiki:Hofstede's_cultural_dimensions_theory}.\\
            
            \item \textit{feature 25}: \textbf{Long-term orientation indicator (LTO)} associates the connection of the past with the current and future actions/challenges. A lower degree of this index (short-term orientation) indicates that traditions are honored and kept \cite{wiki:Hofstede's_cultural_dimensions_theory}.
\\
        \end{itemize}
    \end{itemize}

    \item \textbf{Demographic features} (\textit{features 26-33} in Table \ref{table:features}).\\ These features correspond to traditional migration and population measures obtained from official statistics, either from national censuses or from the population registries.\\
    
    \begin{itemize} 
        \item \textit{feature 26}: annual \textbf{population stocks}, defined as the number of persons having their usual residence in a country in a given year, are gathered both from UN Population Division  \cite{un_pop} (from which only records with "Zero migration" variant were selected) and EUROSTAT \cite{estat_pop}: these two sources often refer to different groups of countries so their mutual integration allowed to cover most of the countries of the dataset. Where both measurements were available for the same country, both were reported.
        The two sources refer to different methodologies, since the annual total population measurement is performed on July 1st by UN, while on January 1st by EUROSTAT. However, their $\sim$1 correlation value proves that the two measures, related to the same year, are well compatible and almost interchangeable: indeed missing values related to the former have been replaced with the latter, and vice versa.\\
        
        \item \textit{feature 27} represents annual \textbf{population density}, defined as the ratio between the annual average population and the land area. Therefore, its unit of measure correponds to "persons per square kilometre". Data has been retrieved from ESTAT \cite{estat_pop_dens}.\\
        
        \item \textit{feature 28, 29}: \textbf{absolute number of migrants} (respectively, immigrants and emigrants) per country. Data was taken from ESTAT \cite{estat_total_imm, estat_total_em} and from UN datasets on flows (see below \textit{feature 32}) selecting, from these latters, records having "Total" as country (respectively, origin and destination country). \\
        
        \item \textit{features 30, 31} indicate quinquennial \textbf{NET migration} and \textbf{NET migration rate} of each country. The former is the difference between the number of immigrants and the number of emigrants in a given area during the reference year, while the latter is defined as the NET migration per 1,000 persons and so it indicates the contribution of migration to the overall level of population change. 
        A positive value for them indicates that there are more migrants entering than leaving a country (NET immigration), while a negative one means that emigrants are more than immigrants (NET emigration).\\
        Values have been taken from UN Population Division \cite{net_migr, net_rate}: note that they apply also for EUROSTAT countries, and they have been widely used in literature in combination with them, even if NET migration rate calculation is based on midyear population (as required by the standard UN methodology).
        \\
        
        \item \textit{feature 32}: yearly \textbf{migration flows} for each pair of countries are defined as the number of people that have moved the country (i.e. that changed residence). Unlike a static stock measure, flow data are dynamic, summarising movements over defined period and consequently allow for a better understanding of past patterns and the prediction of future trends \cite{Abel2016}.
        \\Both EUROSTAT and UN divide migration flows into three categories: by residence \cite{UNInflowRes, UNOutflowRes, ESTATOutflowRes, ESTATInflowRes}, by citizenship \cite{UNInflowCit, ESTATInflowCit} and by country of birth \cite{UNInflowBirth, ESTATInflowsBirth}. This is true in EUROSTAT for both inflows and outflows, while in UN only for inflows, as UN outflows exist only by residence. For our purposes, however, we selected EUROSTAT outflows only by residence, since the ones by citizenship and by country of birth cannot properly be defined "flows", having missing destination country.\\
        
        \item \textit{feature 33}: quinquennial \textbf{migration stocks} for each pair of countries consist in the absolute number of migrants residing in the destination country at given time. Data is obtained from UN \cite{stocks} and includes stocks by sex and age.\\
    \end{itemize}
    
\end{itemize}


\section{Methods}\label{section:methods}
The entire work 
was performed in Python 3.8 language, with the aid of Jupyter software\footnote{\url{https://jupyter.org/}}. \\The initial phase consisted in \textbf{data collection and acquisition}, starting from the exploration of open source portals and proceeding with data selection and download. Initially, only migration flows data were imported.\\

Then a \textbf{pre-processing} phase started, where we carried out data understanding, cleaning and preparation. This has been managed by defining some functions that automatically clean and prepare source datasets. Here our data was subjected to various computational standard processes (such as outliers detection, duplicates handling, uniforming notation, etc…). Some of the operations that have been performed at this level included the selection of task-relevant data (detection of country-to-country valid records, aggregation removal, and non-bilateral flows elimination).\\

\textbf{Data transformation} phase was fundamental to reshape the data in order to resemble the final structure (previously established by our design choices) so that to have a huge matrix with pairs of countries as rows. Concretely, this meant converting, grouping, and unstacking records of source datasets in order to transform them in features (columns).
We continued on shaping this framework by working on indexing: to obtain the dataset index we described in Section \ref{section:features_descr}, duplicates of pairs of countries where not admissible. For this reason, specifically with respect to EUROSTAT flows, we established a priority for selection of pairs: the union of keys (pairs) was taken firstly selecting migration by citizenship, then by residence, and lastly by country of birth.\\

The following step was \textbf{data integration} were we collected, included and computed all other indicators. Geographic and interdisciplinary features related to single countries (\textit{5-25} in Table \ref{table:features}) have been processed in a separate dataset since, neither containing demographic data nor information about couples of countries, it can be reused in different contexts where needed. This \texttt{countries.csv} dataset has undergone the same pre-processing pipeline, but not the trasformation one, since it has its own structure and design: it was then merged\footnote{resembling a SQL left outer join} with the MIMI prototype previously obtained (already structured according to our needs) by matching both countries of origin and destination.\\

Finally the latest features (\textit{2-4} and demographic \textit{26-33}, in Table \ref{table:features}) were integrated by computing them or following the previously described merging process, matching single countries or pairs when needed. Once integration has been completed, it has been helpful to check data semantic and statistics of the resulting dataset and make some random inspections in order to verify the need for a further cleaning step.\\ 

The final \textbf{data quality assessment} phase was one of the longest and most delicate, since many values were missing and this could have had a negative impact on the quality of the desired resulting knowledge. They have been integrated from additional sources reported, for each feature, in Section \ref{section:features_descr}.\\


\section{Usage notes} \label{section:results}
In this section our focus is on documenting and describing salient patterns in distributions and correlations of data. We do not seek to provide causal analyses, nor do we want to imply causal relationships at this stage: however we believe it can be useful to analyze the obtained numerical results since they may guide possible future research and led to some interesting progress in human mobility studies.\\

Unless otherwise specified, correlation values have been computed as simple Pearson's correlation \cite{scipy_pearson}, measuring the linear relationship between two variables: values of -1 or +1 imply an exact linear relationship, while 0 implies no correlation. P-values have been computed in order to confirm of refute the relevance of each correlation value: results are indicated in heatmaps with a number of asterisks proportional to the relevance obtained. \\

\begin{tabular}{lll}
    no asterisks & no relevance & p-value $\geq$ 0.5 \\
    * & little relevance & 0.1 $\leq$ p-value < 0.5 \\
    ** & medium relevance & 0.01 $\leq$ p-value < 0.1 \\
    *** & high relevance & p-value < 0.01 \\
\end{tabular}
\vspace{0.5cm}

When no asterisks are reported for all the values in the matrix, all the correlations computed are highly relevant, meaning p-values always below the threshold of 0.01.\\

\subsection{Data statistics, distributions and correlations.} \label{section:corr}
In this section we provide some practical examples of how to explore the data. A Github repository
(\href{https://bit.ly/GitHubMIMI}{bit.ly/GitHubMIMI}) contains source code of examples of how to read and analyze the data. All the examples are provided in Python 3 language.\\

\begin{figure}
    \centering
    \includegraphics[width = 16 cm]{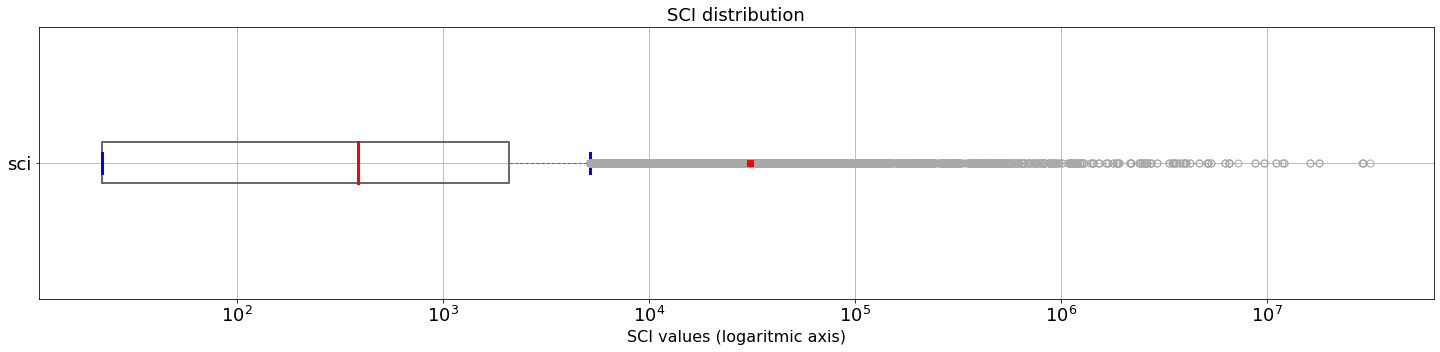}
    \caption{Boxplot of SCI distribution. \label{sci_box}}
    \vspace{1cm}
    
    \includegraphics[width = 16 cm]{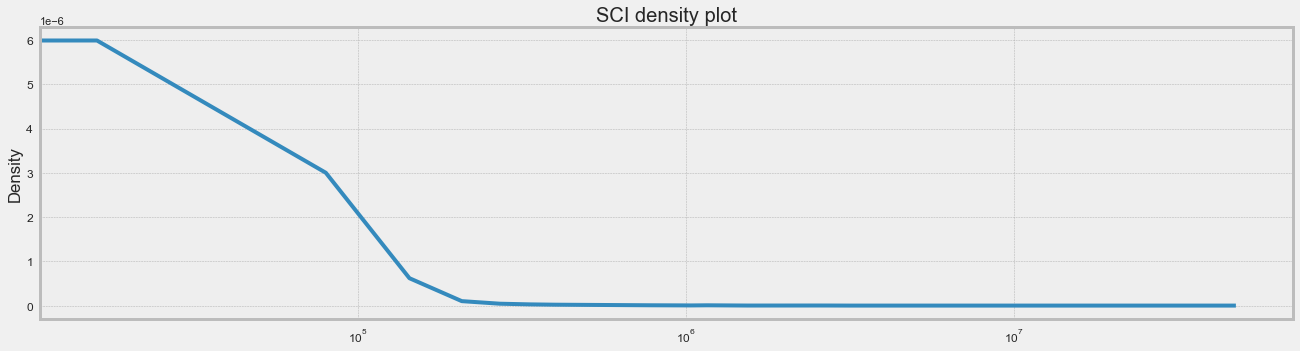}
    \caption{Density plot of SCI with logarithmic x axis. It shows a strongly  right-skewed distribution, meaning that the smallest values of the indicator are the most frequent. \label{sci_density}}
    \vspace{1cm}

    \includegraphics[width = 16 cm]{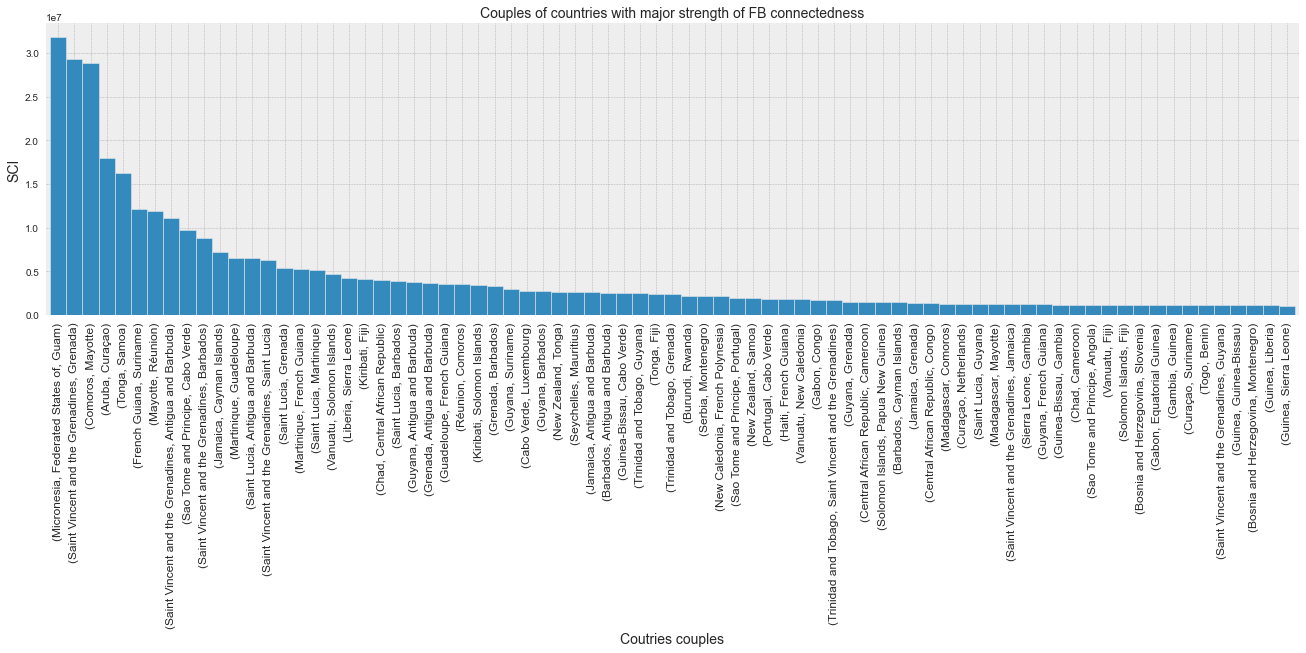}
    \caption{Sample of the highest value of SCI, over 99 quantile. It displays countries pairs with the highest strength of Facebook connectivity. \label{sci_example}}
    \vspace{1cm}

\end{figure}

\begin{figure}
    \centering
    \includegraphics[width = 9 cm]{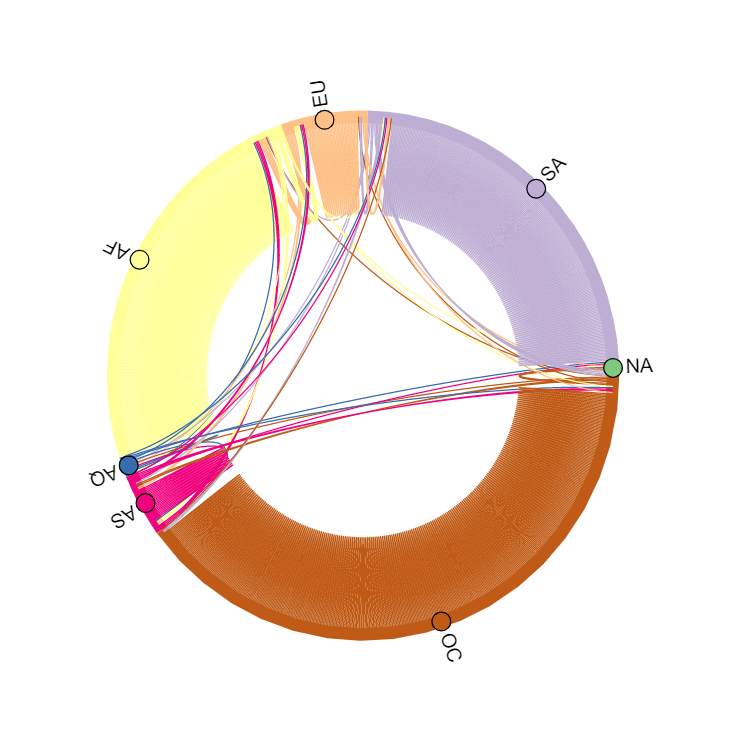}
    \caption{Facebook strength of connectedness among continents (averaged aggregation of SCI for each couple of countries in the continent). Very high values of connectivity can be noticed in Oceania, Africa and South America. Intra-continental connections are much stronger than inter-continental connections, confirming that the intensity of friendship links is strongly declining in geographic distance \cite{Bailey2018}. \label{sci}}
\end{figure}

Figures \ref{sci_box}, \ref{sci_density}, \ref{sci_example} and \ref{sci} show some interesting insights about the distribution of values and the top coverage on global scale of Social Connectedness Index. \\

\begin{figure}
    \centering
    \includegraphics[width = 8 cm]{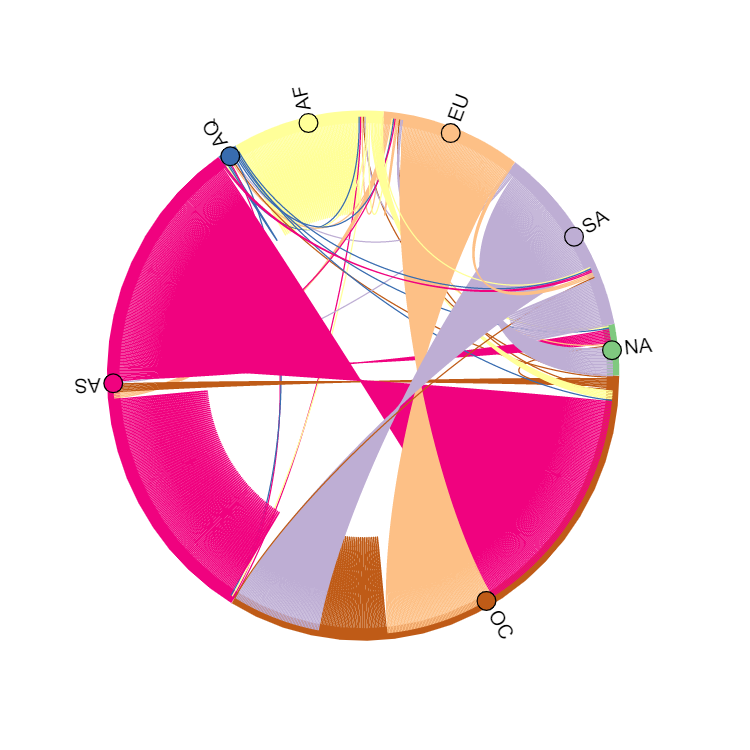}
    \includegraphics[width = 8 cm]{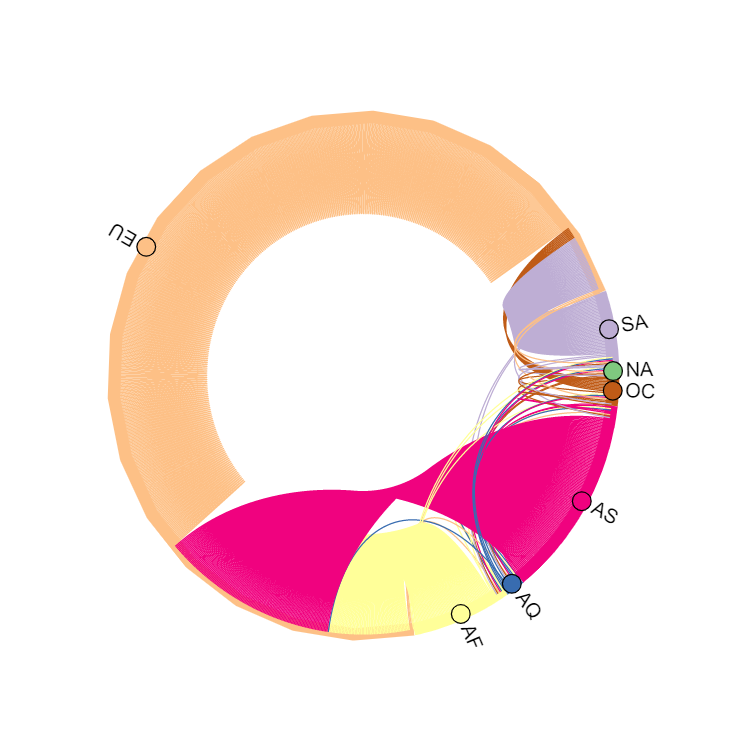}
    \caption{Inter-continental migration flows from UN (left) and EUROSTAT (right) in 2010. We point out a strong intra-continental mobility for Asia and Europe but also relevant immigration trends in Oceania and Europe. In contrast, Asia and South America are subject to almost only emigration. \label{2010}}
\end{figure}

\begin{figure}
    \centering
    \includegraphics[width = 8 cm]{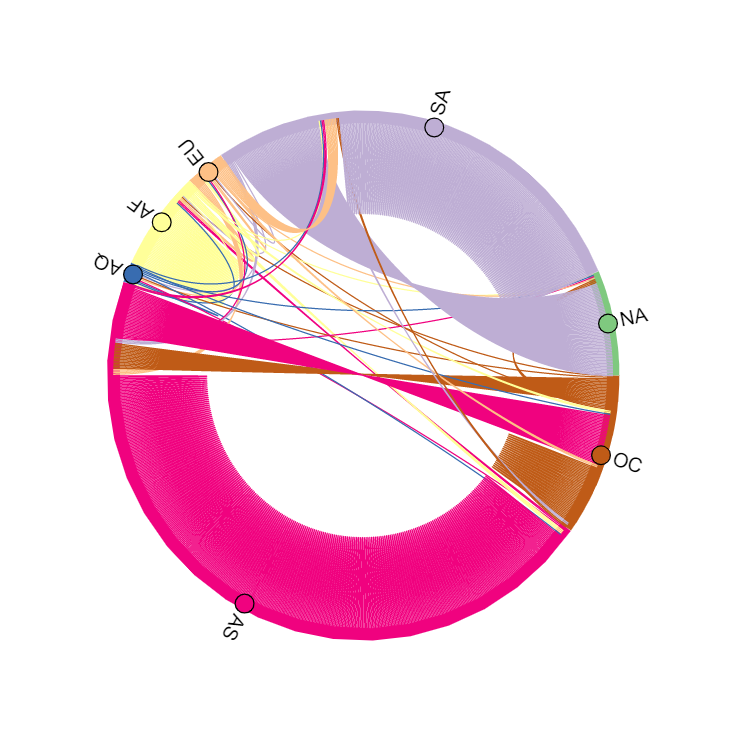}
    \includegraphics[width = 8 cm]{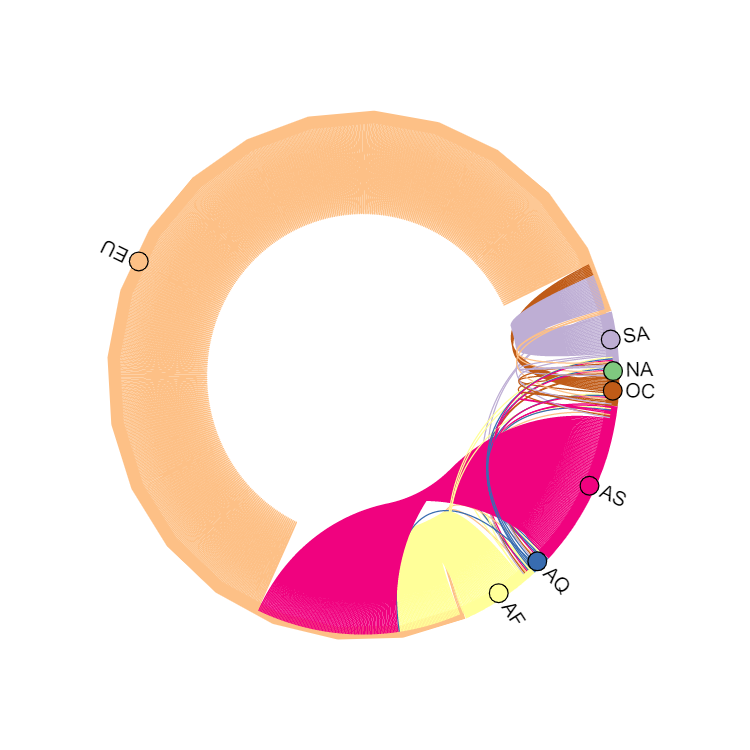}
    \caption{Inter-continental migration flows from UN (left) and EUROSTAT (right) in 2014 \label{2014}. Asia and South America remain continents with a strong emigration, bound for the same destinations as in previous years and which appears even to increase in inter-continental trends.}
\end{figure}

\begin{figure}
    \centering
    \includegraphics[width = 8 cm]{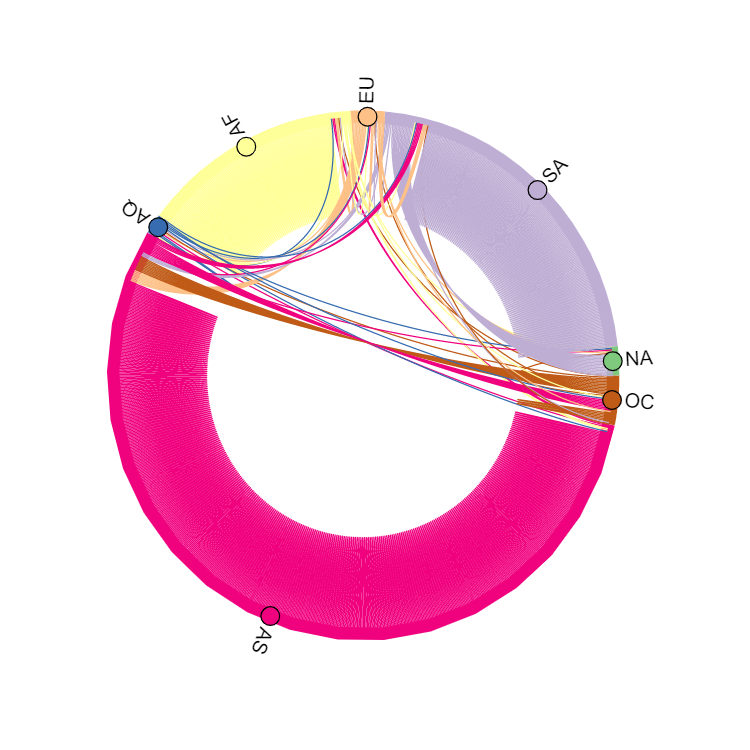}
    \includegraphics[width = 8 cm]{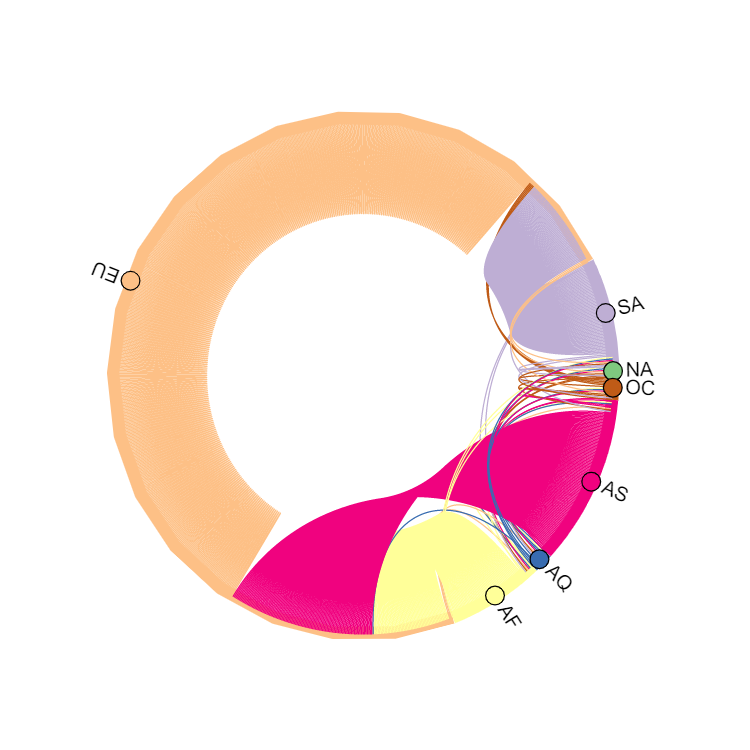}
    \caption{Inter-continental migration flows from UN (left) and EUROSTAT (right) in 2018 is declining with respect to previous years: Oceania experiences far fewer incoming migrants as well as Europe with outgoing ones.\label{2018}}
\end{figure}

\begin{figure}
    \centering
    \includegraphics[width=10cm]{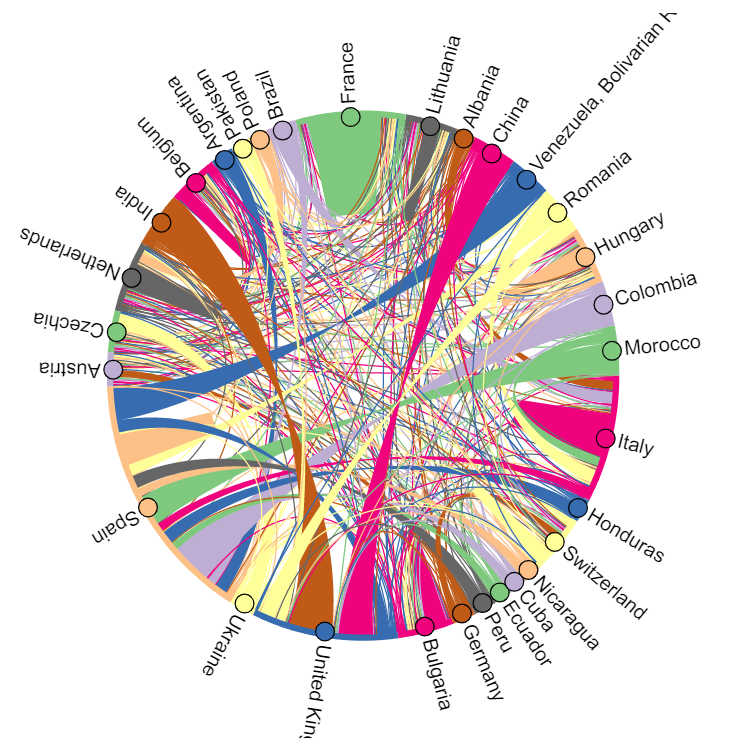}
    \caption{EUROSTAT bilateral migration flows in the most recent year available (2019): pairs of countries with the highest numbers of migrants sharing.}
    \label{fig:ESTAT_2019_flow_chord}
\end{figure}

\begin{figure}
    \centering
    \includegraphics[width=10cm]{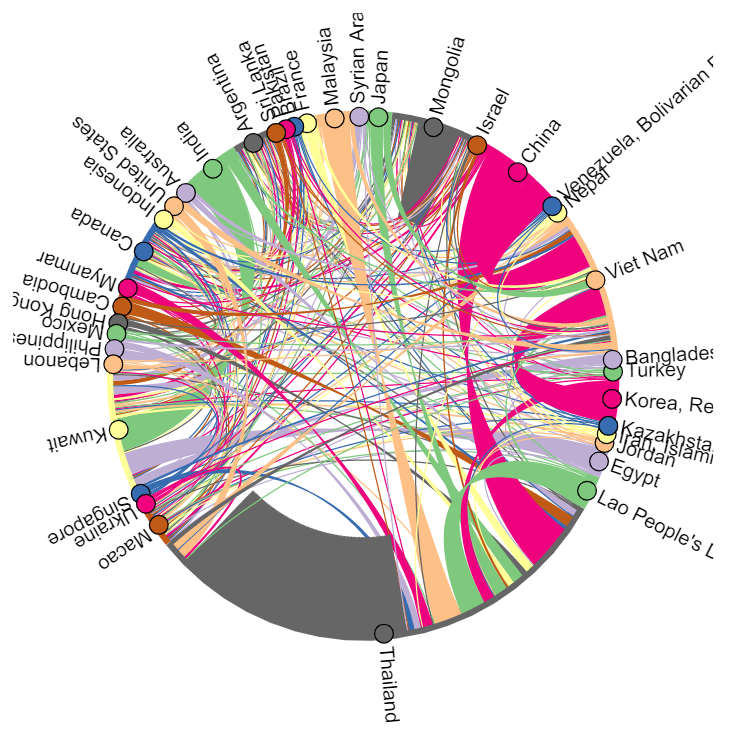}
    \caption{UN bilateral migration flows in the most recent year available (2020): pairs of countries with the highest numbers of migrants sharing.} 
    \label{fig:UN_2020_flow_chord}
\end{figure}

\begin{figure}
    \centering
    \includegraphics[width=10cm]{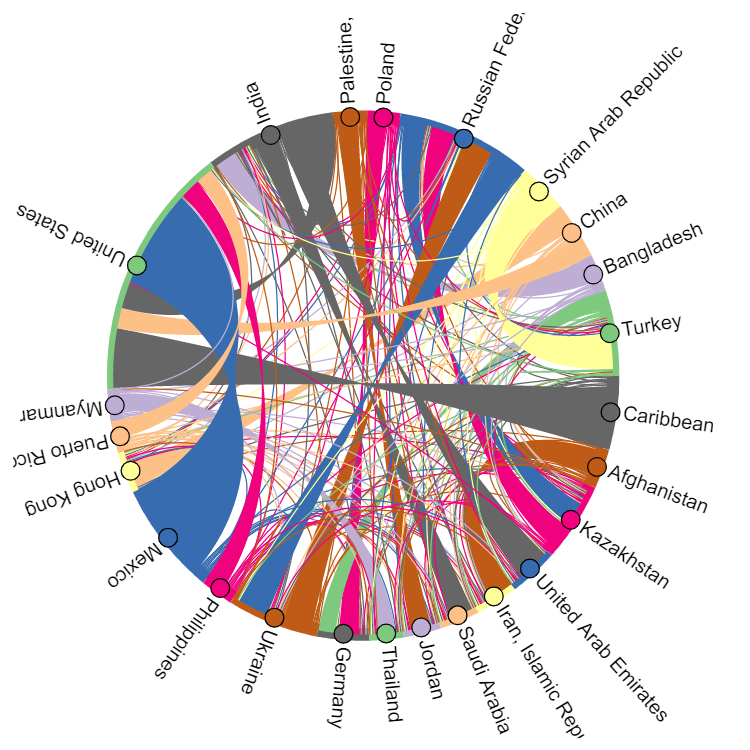}
    \caption{UN bilateral migration stocks in the most recent year available (2020): pairs of countries with the highest numbers of permanent residing migrants.}
    \label{fig:UN_2020_stock_chord}
\end{figure}

Chord diagrams illustrate bidirectional international mobility between pairs of continents in Figures \ref{2010}, \ref{2014} and \ref{2018} (where values are aggregated as total sum of migration flows shared by pairs of countries in the given continent), and between pairs countries in Figures \ref{fig:ESTAT_2019_flow_chord}, \ref{fig:UN_2020_flow_chord} and \ref{fig:UN_2020_stock_chord}.\\

\begin{figure}
    \centering
    \includegraphics[width = 14 cm]{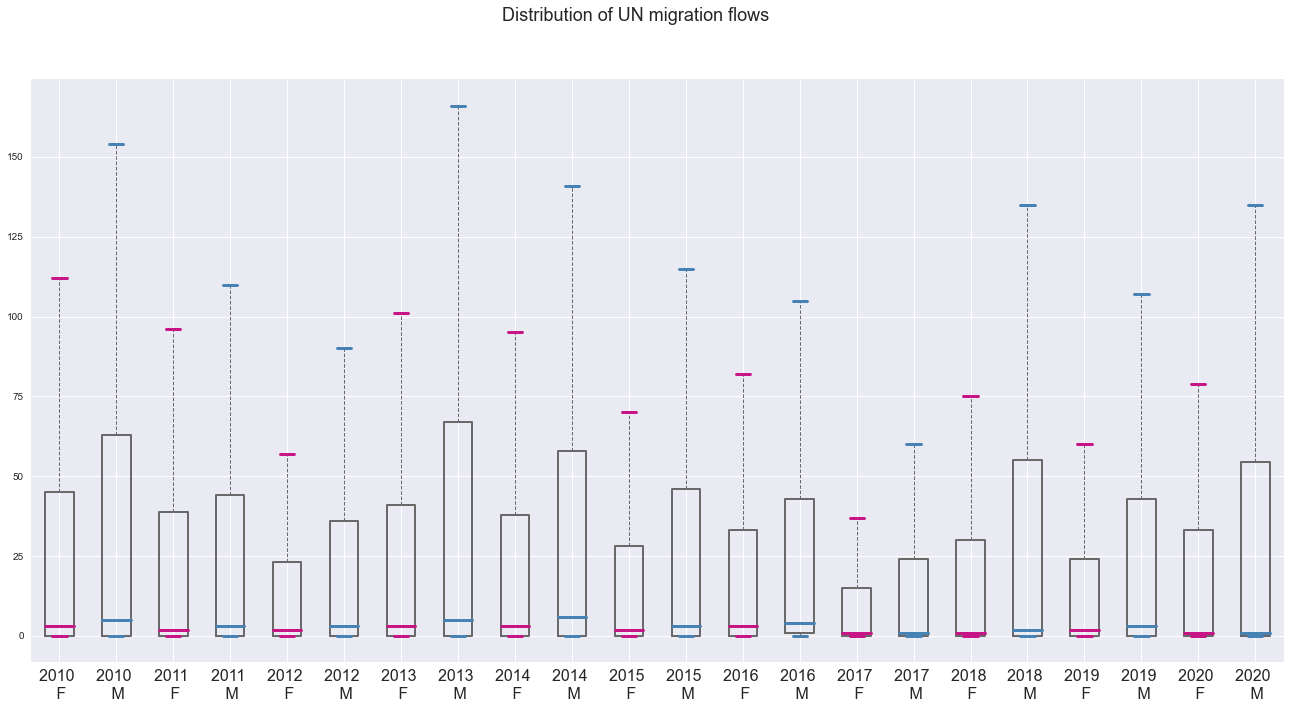}
    \caption{Distribution of migration flows from UN. The increasing trend encountered in the previous chart is not present for these distributions, where instead it is possible to notice a regularity in the behavior over time: a gradual descent takes a few years (which ends coincide with the drops in the previous plot) and then have a sudden peak of ascent. The discrepancy between male and female migration is sharper.\label{UN_flows_box}}
\end{figure}

Despite the impact of COVID-19 pandemic on international human mobility, mostly related to travel restrictions and "stay-at-home" measures which reduced internal movements within a country \cite{wmr2022}, Figures \ref{fig:UN_2020_flow_chord} and \ref{UN_flows_box} confirm that the numbers in migration flows statistics did not suffer. However, a consistent flow of returners can be noticed for Thailand, probably due to COVID-19 itself, since in 2020 the pandemic prompted the return of hundreds of thousands of migrants to their countries of origin \cite{migr_data_portal}.\\
Regarding migration stocks in Figure \ref{fig:UN_2020_stock_chord}, the impact of COVID-19 on the global population of international migrants is difficult to assess, since the latest available data refers mid-2020, fairly early in the pandemic. However, it is estimated that the pandemic may have reduced the growth in the stock of international migrants by around two million \cite{wmr2022, mrs2020}.
\\ Moreover, almost all these pairs of countries are included in the "top 20 international migration country-to-country corridors, 2020" list in the World Migration Report 2022 \cite{wmr2022}, (e.g. Mexico – United States, Syria – Turkey, India – Saudi Arabia, United Arab Emirates and United States, Afghanistan – Iran, Myanmar – Thailand), meaning that the greatest communities of permanently residing migrants in a host country have developed over years for safety reasons.\\

\begin{figure}
    \centering
    \includegraphics[width = 14 cm]{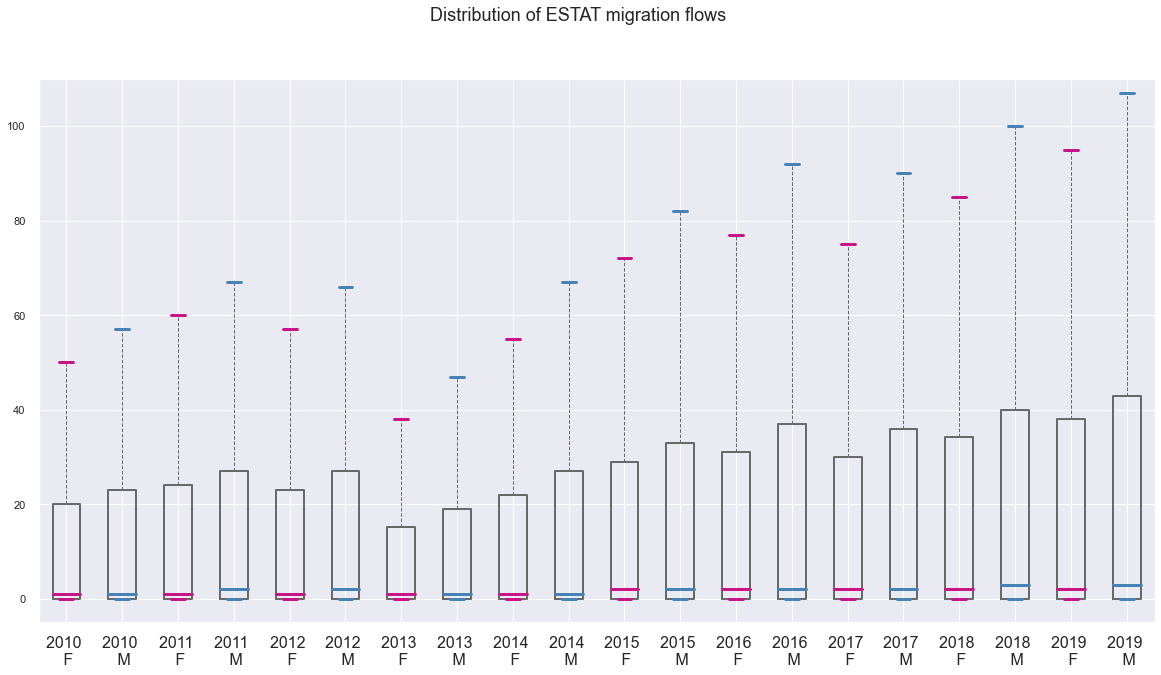}
    \caption{Distribution of migration flows from EUROSTAT. Male migration is always higher than female migration for each annual measurement, while the general trend over time is a slight increase of the migration phenomenon. Two drops in the progressive grown of values can be identified, corresponding to triennium 2012-2014 and to 2017. \label{ESTAT_flows_box}}
\end{figure}

\begin{figure}
    \centering
    \includegraphics[width = 8 cm]{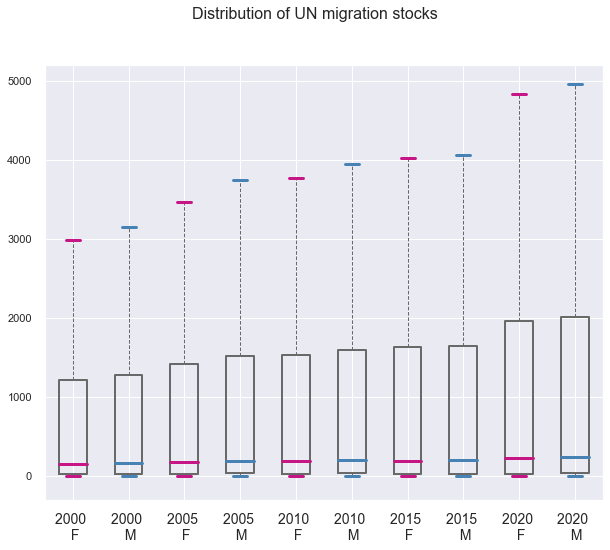}
    \caption{Distribution of migration stocks. The five year measurement prevents you from having a more detailed look as it was for the flows: nevertheless, an increase in the general trend over years is quite evident. \label{UN_stocks_box}}
\end{figure}

Boxplots in Figures \ref{ESTAT_flows_box}, \ref{UN_flows_box} and \ref{UN_stocks_box} display the statistical distribution of migration flows and stocks values over the years, divided by sex. Increasing trends and regular patterns over time are well recognizable from the timeseries data plotted, as well as the statistics evidence on male migration that reveals largest numbers with respect to female one (about gender dimensions on human mobility refer to \cite{wmr2018}).\\ 

\begin{figure}
    \centering
    \includegraphics[width = 14 cm]{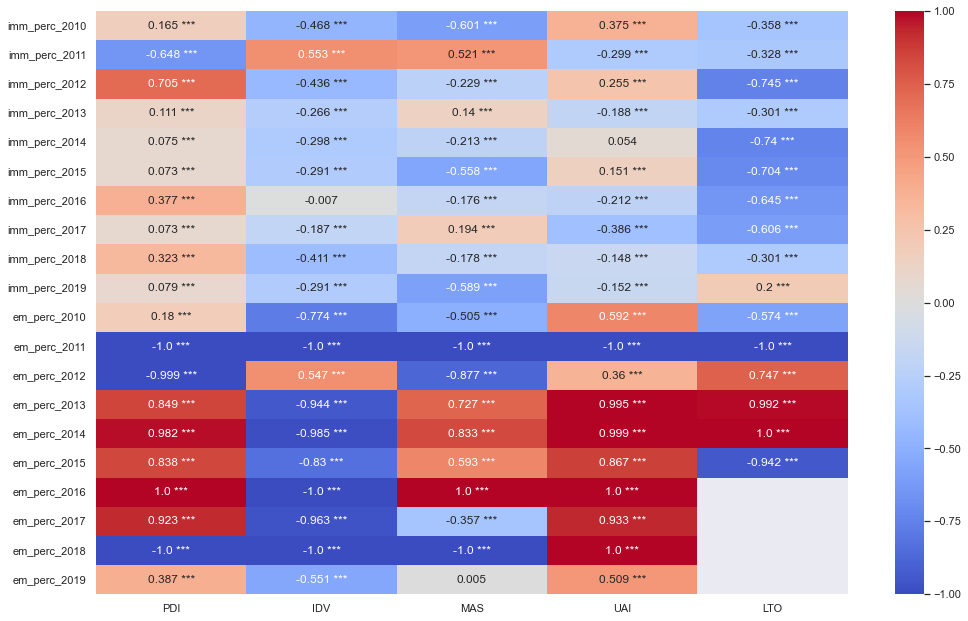}
    \caption{Correlation between immigrants / emigrants over years and cultural indicators of a country. Absolute numbers of total migrants have been divided by the annual total population of the country. Strong positive values are indicated in red while strong negative values in blue. Asterisks indicates the relevance of the p-values obtained, as described in Section \ref{section:results}. \label{cult_indicators_corr}}
    \vspace{1cm}
\end{figure}

\begin{figure}
    \centering
    \includegraphics[width = 14 cm]{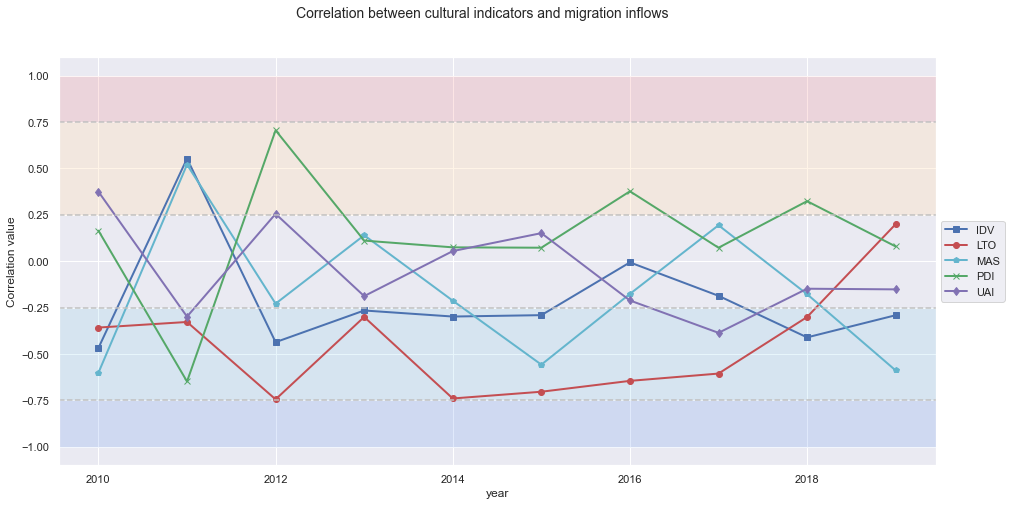}
    \caption{Distribution of correlation between total immigration and cultural indicators. Immigrants for each year are expressed as ratio with respect to the total population of the country for the same year. \label{cult_indicators_imm_hist}}
    \vspace{1cm}
\end{figure}

\begin{figure}
    \centering
    \includegraphics[width = 14 cm]{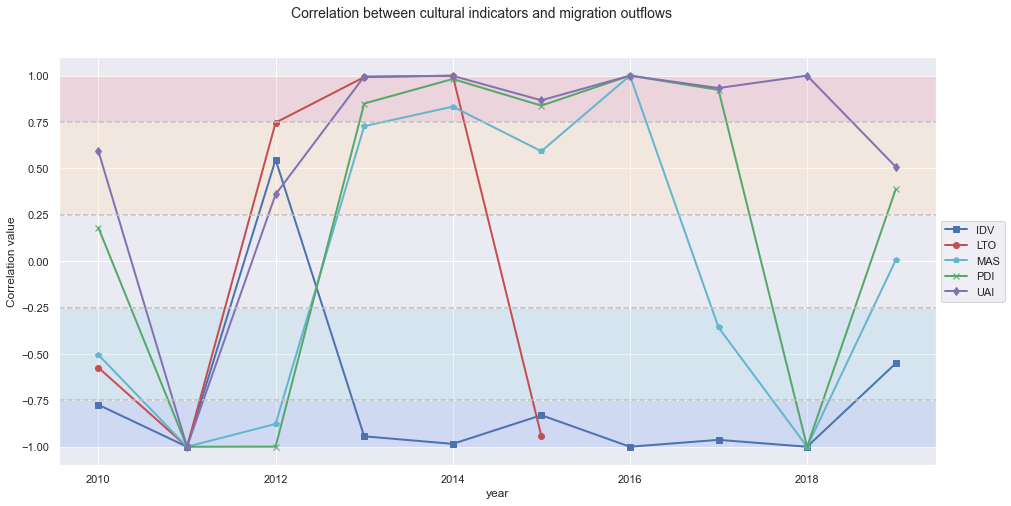}
    \caption{Distribution of correlation between total emigration and cultural indicators. Emigrants for each year are expressed as ratio with respect to the total population of the country for the same year. \label{cult_indicators_em_hist}}
    \vspace{1cm}
\end{figure}

Heatmap in Figure \ref{cult_indicators_corr} shows correlations between the computed ratio of total migrants and total population of a country and its cultural indicators, while Figures \ref{cult_indicators_imm_hist} and \ref{cult_indicators_em_hist} correspond to the outcome of the division of that heatmap in immigration and emigration with the mapping of annual correlation values in a bi-dimensional plane. 

Values of almost all indicators seem to initially lie mostly in the upper zone of the plane, showing a quite strong positive correlation with emigration, until some breakpoint years occur and the correlation value becomes henceforth highly negative.
This radical change in trend cannot yet be supported and explained  by a causal relation, so we limit ourselves to report its behavior.\\
Concerning correlation related to immigration, they lie on the middle region of the plot, quite far from the range in the upper and lower extremities, therefore assuming less polarized values. Besides, there are no trend reversal for it. \\

\begin{figure}
    \centering
    \includegraphics[width = 12 cm]{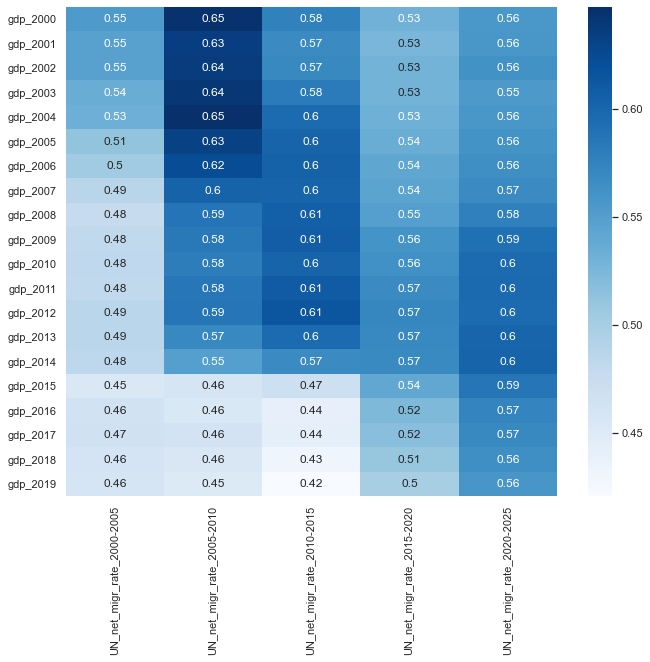}
    \caption{Correlation matrix between annual GDP per capita and five-year NET migration rate of a country. \label{NET_rate_GDP}}
\end{figure}

Correlation between NET migration rate and GDP of a country shown in Figure \ref{NET_rate_GDP} confirms the existing relation, well documented in literature, between these two variables. Correlation is always positive, meaning that countries with high GDP face a NET immigration trend and so confirming that high per capita income are conducive to mobility \cite{bell2015}. Specifically, human mobility is influenced by GDP values up to more than 10 years back.

\begin{figure}
    \centering
    \includegraphics[width = 12 cm]{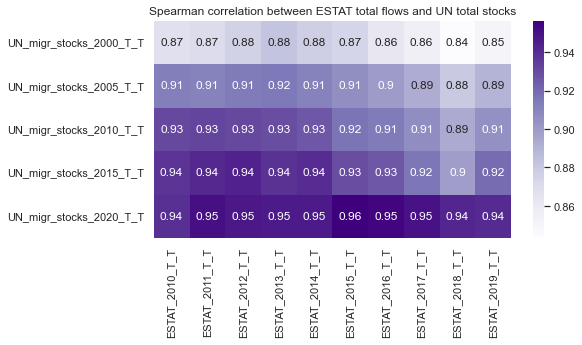}
    \includegraphics[width = 12 cm]{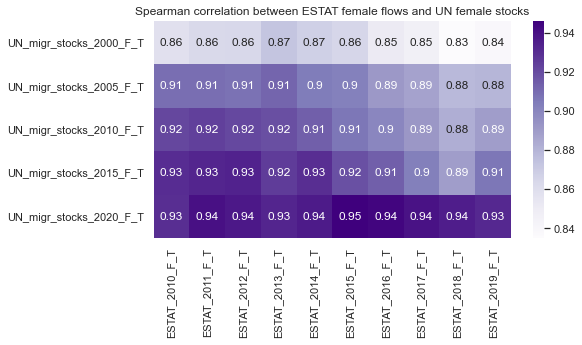}
    \includegraphics[width = 12 cm]{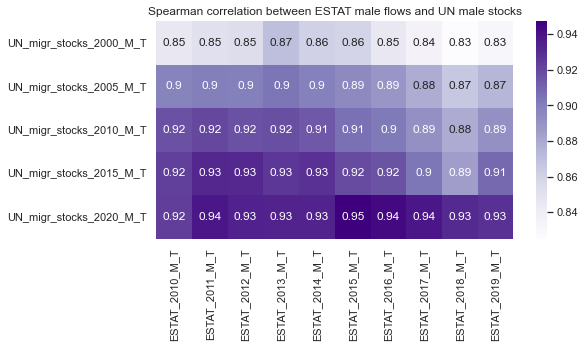}
    \vspace{0.5cm}
    \caption{Spearman correlation between migration flows and stocks, divided by sex. \label{ESTAT_flows_stocks}}
\end{figure}

Heatmaps in Figure \ref{ESTAT_flows_stocks} illustrates the trends in Spearman correlations over years between EUROSTAT migrations flows and UN migration stocks. 
Although the existing correlation between stocks at a given time \textit{t} and flows relative to previous years is self-evident (as those same flows will be included in the total counting of stocks), it is interesting to notice that quite strong positive correlations also propagate forward in time: this could mean that the higher the stock count at a given time \textit{t}, the more migration flows will be shared by the pair of countries.

\begin{figure}
    \centering
    \includegraphics[width = 14 cm]{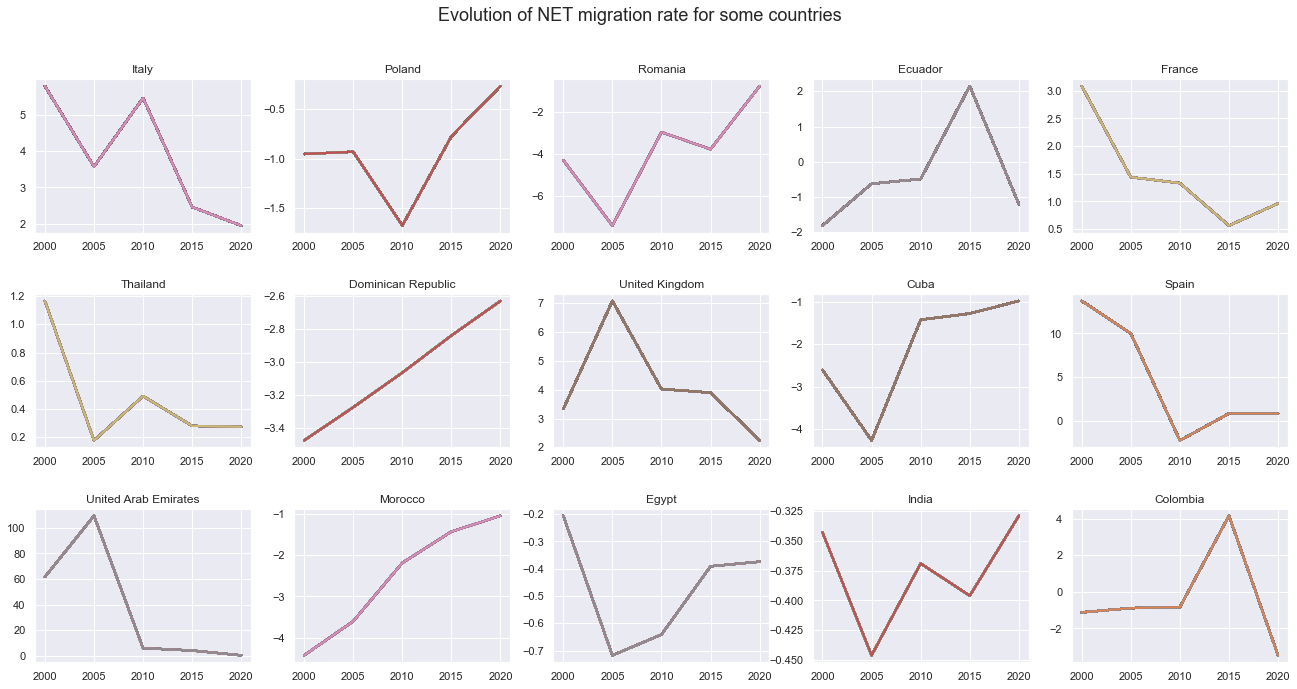}
    \caption{Evolution of five-year NET migration rate over time, for a sample of countries. \label{NET_rate}}
\end{figure}

Finally, Figure \ref{NET_rate} explores the changes in trend of NET migration rate for a small sample of countries.


\clearpage

\appendix
\section*{Abbreviations}{\label{abbr}
The following abbreviations are used in this work:\\
\noindent 
\begin{tabular}{@{}ll}
GDP & Gross Domestic Product\\
PPP & Purchasing Power Parity\\
SCI & Social Connectedness Index\\
UN & United Nations\\
\end{tabular}}

\section*{Funding Statement}
This research was funded by HumMingBird H2020 Program grant number 870661 and SoBigData grant number 871042.

\section*{Data availability}
The MIMI dataset v1.0 presented in this study is openly available on Zenodo at the following link: \href{https://doi.org/10.5281/zenodo.6360651}{doi.org/10.5281/zenodo.6360651}.

\bibliographystyle{abbrv}
\bibliography{mimibib}

\begin{thebibliography}{10}

\bibitem{Abel2016}
G.~J. Abel.
\newblock Estimates of global bilateral migration flows by gender between 1960
  and 2015, 2016.

\bibitem{alexander2022}
M.~Alexander, K.~Polimis, and E.~Zagheni.
\newblock Combining social media and survey data to nowcast migrant stocks in
  the united states.
\newblock {\em Population Research and Policy Review}, 41, 02 2022.

\bibitem{ieee2019}
I.~Annamoradnejad, M.~Fazli, J.~Habibi, and S.~Tavakoli.
\newblock Cross-cultural studies using social networks data.
\newblock {\em IEEE Transactions on Computational Social Systems}, PP:1--10, 06
  2019.

\bibitem{Bailey2018}
M.~Bailey, R.~Cao, T.~Kuchler, J.~Stroebel, and A.~Wong.
\newblock Social connectedness: Measurement, determinants, and effects.
\newblock {\em Journal of Economic Perspectives}, 32(3):259--80, August 2018.

\bibitem{bailey2020determinants}
M.~Bailey, D.~Johnston, T.~Kuchler, D.~Russel, B.~State, and J.~Stroebel.
\newblock The determinants of social connectedness in europe.
\newblock In {\em Social Informatics}. Springer International Publishing, 2020.

\bibitem{bell2015}
M.~Bell, E.~Charles-Edwards, P.~Ueffing, J.~Stillwell, M.~Kupiszewski, and
  D.~Kupiszewska.
\newblock Internal migration and development: Comparing migration intensities
  around the world.
\newblock {\em Population and Development Review}, 41, 03 2015.

\bibitem{JRC127369}
B.~Claudio, G.-B. Sara, I.~Stefano, M.~Umberto, S.~Francesco, and S.~Spyridon.
\newblock Data innovation in demography, migration and human mobility.
\newblock 2022.
\newblock DOI:
  \href{https://www.researchgate.net/publication/358695486_Data_Innovation_in_Demography_Migration_and_Human_Mobility}{10.2760/958409}.

\bibitem{datahub_religion}
DataHub.
\newblock World religion projections.
\newblock \url{https://datahub.io/sagargg/world-religion-projections}.
\newblock [Online; accessed December 2021.].

\bibitem{cultdim2}
data.world.
\newblock Geerthofstedeculturaldimension.
\newblock \url{https://bit.ly/CultDimDataset}.
\newblock [Online; accessed December 2021.].

\bibitem{delfava2019}
E.~Del~Fava, A.~Wiśniowski, and E.~Zagheni.
\newblock Modelling international migration flows by integrating multiple data
  sources.
\newblock 11 2019.

\bibitem{DSPL}
G.~DSPL.
\newblock Countries dataset.
\newblock
  \url{https://developers.google.com/public-data/docs/canonical/countries_csv}.
\newblock [Online; accessed December 2021.].

\bibitem{estat_total_em}
Eurostat.
\newblock Emigration by age and sex.
\newblock \url{https://bit.ly/ESTATEmig}.
\newblock [Online; accessed December 2021.].

\bibitem{ESTATOutflowRes}
Eurostat.
\newblock Emigration by age group, sex and country of next usual residence.
\newblock \url{https://bit.ly/ESTATOutflowRes}.
\newblock [Online; accessed December 2021.].

\bibitem{estat_total_imm}
Eurostat.
\newblock Immigration by age and sex.
\newblock \url{https://bit.ly/ESTATImmig}.
\newblock [Online; accessed December 2021.].

\bibitem{ESTATInflowCit}
Eurostat.
\newblock Immigration by age group, sex and citizenship.
\newblock \url{https://bit.ly/ESTATInflowCit}.
\newblock [Online; accessed December 2021.].

\bibitem{ESTATInflowsBirth}
Eurostat.
\newblock Immigration by age group, sex and country of birth.
\newblock \url{https://bit.ly/ESTATInflowsBirth}.
\newblock [Online; accessed December 2021.].

\bibitem{ESTATInflowRes}
Eurostat.
\newblock Immigration by age group, sex and country of previous residence.
\newblock \url{https://bit.ly/ESTATInflowRes}.
\newblock [Online; accessed December 2021.].

\bibitem{estat_pop_dens}
Eurostat.
\newblock Population density.
\newblock \url{https://bit.ly/ESTATPopDens}.
\newblock [Online; accessed December 2021.].

\bibitem{estat_pop}
Eurostat.
\newblock Population on 1 january.
\newblock \url{https://bit.ly/ESTATPop}.
\newblock [Online; accessed December 2021.].

\bibitem{github2017}
GitHub.
\newblock "bordering-countries" github repository, "neighbors.csv" dataset.
\newblock \url{https://github.com/evpu/Bordering-Countries}, 2017.
\newblock [Online; accessed December 2021.].

\bibitem{goglia2022}
D.~Goglia.
\newblock {Multi-aspect Integrated Migration Indicators (MIMI) dataset}, March
  2022.

\bibitem{Hofstede1980}
G.~Hofstede.
\newblock Culture's consequences: International differences in work related
  valuese.
\newblock 1980.

\bibitem{ISAN2021}
ISAN.
\newblock List of iso 3166 country codes.
\newblock \url{https://bit.ly/ISANmetadata}.
\newblock [Online; accessed December 2021.].

\bibitem{cultdim1}
A.~Kaasa.
\newblock Ess/evs-based indicators of cultural dimensions.
\newblock \url{https://bit.ly/CultDim}.
\newblock [Online; accessed December 2021.].

\bibitem{KAASA2016231}
A.~Kaasa, M.~Vadi, and U.~Varblane.
\newblock A new dataset of cultural distances for european countries and
  regions.
\newblock {\em Research in International Business and Finance}, 37:231--241,
  2016.

\bibitem{Kaasa2014}
V.~{Kaasa, Anneli}, {Vadi, Maaja}.
\newblock Regional cultural differences within european countries: Evidence
  from multi-country surveys.
\newblock {\em Management International Review}, 54:825--852, 2014.

\bibitem{Karney2013}
C.~F.~F. Karney.
\newblock Algorithms for geodesics.
\newblock {\em Journal of Geodesy}, 87:43–55, 2013.

\bibitem{Lazarsfeld1954}
P.~Lazarsfeld and R.~K. Merton.
\newblock Friendship as a social process: A substantive and methodological
  analysis.
\newblock {\em Freedom and Control in Modern Society}, 1954.

\bibitem{mrs2020}
M.~McAuliffe.
\newblock {\em Immobility as the ultimate migration disrupter}.
\newblock Number~64. Migration Research Series, 2020.

\bibitem{wmr2018}
M.~McAuliffe, A.~Kitimbo, A.~Goossens, and A.~Ullah.
\newblock Understanding migration journeys from migrants' perspectives.
\newblock In {\em World Migration Report 2018}. International Organization for
  Migration (IOM), Geneva, 2017.

\bibitem{wmr2022}
M.~McAuliffe and A.~Triandafyllidou.
\newblock Migration and migrants: A global overview.
\newblock In {\em World Migration Report 2022}. International Organization for
  Migration (IOM), Geneva, 2021.

\bibitem{sci2021}
Meta.
\newblock Social connectedness index.
\newblock \url{https://bit.ly/SCIdataset}, 2021.
\newblock [Online; accessed December 2021.].

\bibitem{migr_data_portal}
MigrationDataPortal.
\newblock Migration data in south-eastern asia.
\newblock
  \url{https://www.migrationdataportal.org/regional-data-overview/south-eastern-asia}.
\newblock [Online; accessed March 2022.].

\bibitem{scipy_pearson}
SciPy.
\newblock scipy.stats.pearsonr.
\newblock
  \url{https://docs.scipy.org/doc/scipy/reference/generated/scipy.stats.pearsonr.html#r8c6348c62346-3}.
\newblock [Online; accessed February 2022.].

\bibitem{sirbu2021}
A.~Sîrbu, G.~Andrienko, N.~Andrienko, C.~Boldrini, M.~Conti, F.~Giannotti,
  R.~Guidotti, S.~Bertoli, J.~Kim, C.~I. Muntean, L.~Pappalardo, A.~Passarella,
  D.~Pedreschi, L.~Pollacci, F.~Pratesi, and R.~Sharma.
\newblock Human migration: the big data perspective.
\newblock {\em International Journal of Data Science and Analytics},
  11(4):341--360, 2021.

\bibitem{twb_gdp}
TheWorldBank.
\newblock Gdp per capita, ppp (current international \$).
\newblock \url{https://data.worldbank.org/indicator/NY.GDP.PCAP.PP.CD}.
\newblock [Online; accessed December 2021.].

\bibitem{twb_area}
TheWorldBank.
\newblock Land area (sq. km).
\newblock \url{https://data.worldbank.org/indicator/AG.LND.TOTL.K2}.
\newblock [Online; accessed December 2021.].

\bibitem{UNInflowBirth}
UN.
\newblock Foreign-born population by country/area of birth.
\newblock \url{https://bit.ly/UNInflowBirth}.
\newblock [Online; accessed December 2021.].

\bibitem{stocks}
UN.
\newblock International migrant stock 2020.
\newblock \url{https://bit.ly/UNStocks}.
\newblock [Online; accessed December 2021.].

\bibitem{net_rate}
UN.
\newblock Net migration rate (per 1,000 population).
\newblock \url{https://bit.ly/NETrate}.
\newblock [Online; accessed December 2021.].

\bibitem{net_migr}
UN.
\newblock Net number of migrants, both sexes combined (thousands).
\newblock \url{https://bit.ly/NETmigr}.
\newblock [Online; accessed December 2021.].

\bibitem{UNOutflowRes}
UN.
\newblock Number of emigrating citizens by future country of usual residence
  and sex.
\newblock \url{https://bit.ly/UNOutflowRes}.
\newblock [Online; accessed December 2021.].

\bibitem{UNInflowCit}
UN.
\newblock Number of incoming foreign migrants by country of citizenship and
  sex.
\newblock \url{https://bit.ly/UNInflowCit}.
\newblock [Online; accessed December 2021.].

\bibitem{UNInflowRes}
UN.
\newblock Number of incoming international migrants by previous country of
  usual residence and sex.
\newblock \url{https://bit.ly/UNInflowRes}.
\newblock [Online; accessed December 2021.].

\bibitem{un_pop}
UN.
\newblock Total population, both sexes combined (thousands).
\newblock \url{https://bit.ly/PopStocks}.
\newblock [Online; accessed December 2021.].

\bibitem{wiki:List_of_countries}
Wikipedia.
\newblock List of countries and territories by land borders.
\newblock
  \url{https://en.wikipedia.org/wiki/List_of_countries_and_territories_by_land_borders}.
\newblock [Online; accessed December 2021.].

\bibitem{wiki:lang}
Wikipedia.
\newblock List of official languages by country and territory.
\newblock
  \url{https://en.wikipedia.org/wiki/List_of_official_languages_by_country_and_territory}.
\newblock [Online; accessed December 2021.].

\bibitem{wiki:religion}
Wikipedia.
\newblock Religions by country.
\newblock \url{https://en.wikipedia.org/wiki/Religions_by_country}.
\newblock [Online; accessed December 2021.].

\bibitem{wiki:Hofstede's_cultural_dimensions_theory}
Wikipedia.
\newblock {Hofstede's cultural dimensions theory} --- {W}ikipedia{,} the free
  encyclopedia.
\newblock
  \url{http://en.wikipedia.org/w/index.php?title=Hofstede's\%20cultural\%20dimensions\%20theory&oldid=1074491009},
  2022.
\newblock [Online; accessed 15-March-2022].

\bibitem{wpr}
WorldPopulationReview.
\newblock Facebook users by country 2022.
\newblock
  \url{https://worldpopulationreview.com/country-rankings/facebook-users-by-country}.
\newblock [Online; accessed December 2021.].

\end{thebibliography}

\end{document}